\documentclass[
twocolumn,
prd,tightenlines,showpacs,nofootinbib,
eqsecnum,
amsfonts,amsmath,amssymb]{revtex4}

\usepackage{bm}
\usepackage{graphicx} 
\usepackage{hyperref}
\usepackage{mathrsfs}
\usepackage{multirow}

\newcommand{\ud}{\mathrm{d}}

\newcommand{\calO}{\mathcal{O}}

\newcommand{\ul}[1]{\underline{#1}{}}

\newcommand{\be}[1]{\begin{equation}\label{#1}}
\newcommand{\ee}{\end{equation}}
\newcommand{\bes}[1]{\begin{subequations}\label{#1}\begin{eqnarray}}
\newcommand{\ees}{\end{eqnarray}\end{subequations}}

\usepackage{ulem}
\usepackage[usenames]{color}

\allowdisplaybreaks

\begin{document}

\markboth{Luc Blanchet \& Laura Bernard} {Phenomenology of MOND and
  Gravitational Polarization}

%
%

\title{PHENOMENOLOGY OF MOND AND GRAVITATIONAL
  POLARIZATION\footnote{To appear in the proceedings of the
    $2^{\text{nd}}$ Workshop on Antimatter and Gravity (WAG 2013).}}

\author{Luc \textsc{Blanchet}}\email{blanchet@iap.fr}
\affiliation{$\mathcal{G}\mathbb{R}\varepsilon{\mathbb{C}}\mathcal{O}$
  Institut d'Astrophysique de Paris --- UMR 7095 du CNRS,
  \ Universit\'e Pierre \& Marie Curie, 98\textsuperscript{bis}
  boulevard Arago, 75014 Paris, France}

\author{Laura \textsc{Bernard}}\email{bernard@iap.fr}
\affiliation{$\mathcal{G}\mathbb{R}\varepsilon{\mathbb{C}}\mathcal{O}$
  Institut d'Astrophysique de Paris --- UMR 7095 du CNRS,
  \ Universit\'e Pierre \& Marie Curie, 98\textsuperscript{bis}
  boulevard Arago, 75014 Paris, France}

\date{\today}

\begin{abstract}
The phenomenology of MOND (flat rotation curves of galaxies, baryonic
Tully-Fisher relation, \textit{etc.}) is a basic set of phenomena
relevant to galaxy dynamics and dark matter distribution at galaxy
scales. Still unexplained today, it enjoys a remarkable property,
known as the \textit{dielectric analogy}, which could have
far-reaching implications. In the present paper we discuss this
analogy in the framework of simple non-relativistic models. We show
how a specific form of dark matter, made of two different species of
particles coupled to different Newtonian gravitational potentials,
could permit to interpret in the most natural way the dielectric
analogy of MOND by a mechanism of gravitational
polarization.\keywords{dark matter, modified gravity}
\end{abstract}

\pacs{04.20.-q, 95.35.+d, 95.30.Sf}

\maketitle

\section{Motivations for MOND}
\label{sec:MOND}

The Modified Newtonian Dynamics (MOND) was introduced more than 30
years ago by Milgrom~\cite{Milg1,Milg2,Milg3} as an alternative to
dark matter, designed to explain a variety of phenomena taking place
at the scale of galaxies, which are now collectively referred to as
the phenomenology of MOND (see Refs.~\cite{SandMcG02,FamMcG12} for
reviews). The ability of MOND at reproducing this phenomenology is
astonishing, and it is fair to say that this still represents a
complete mystery today.

\begin{figure}[htb]
\vspace{-1cm} \centerline{\includegraphics[width=9.5cm]{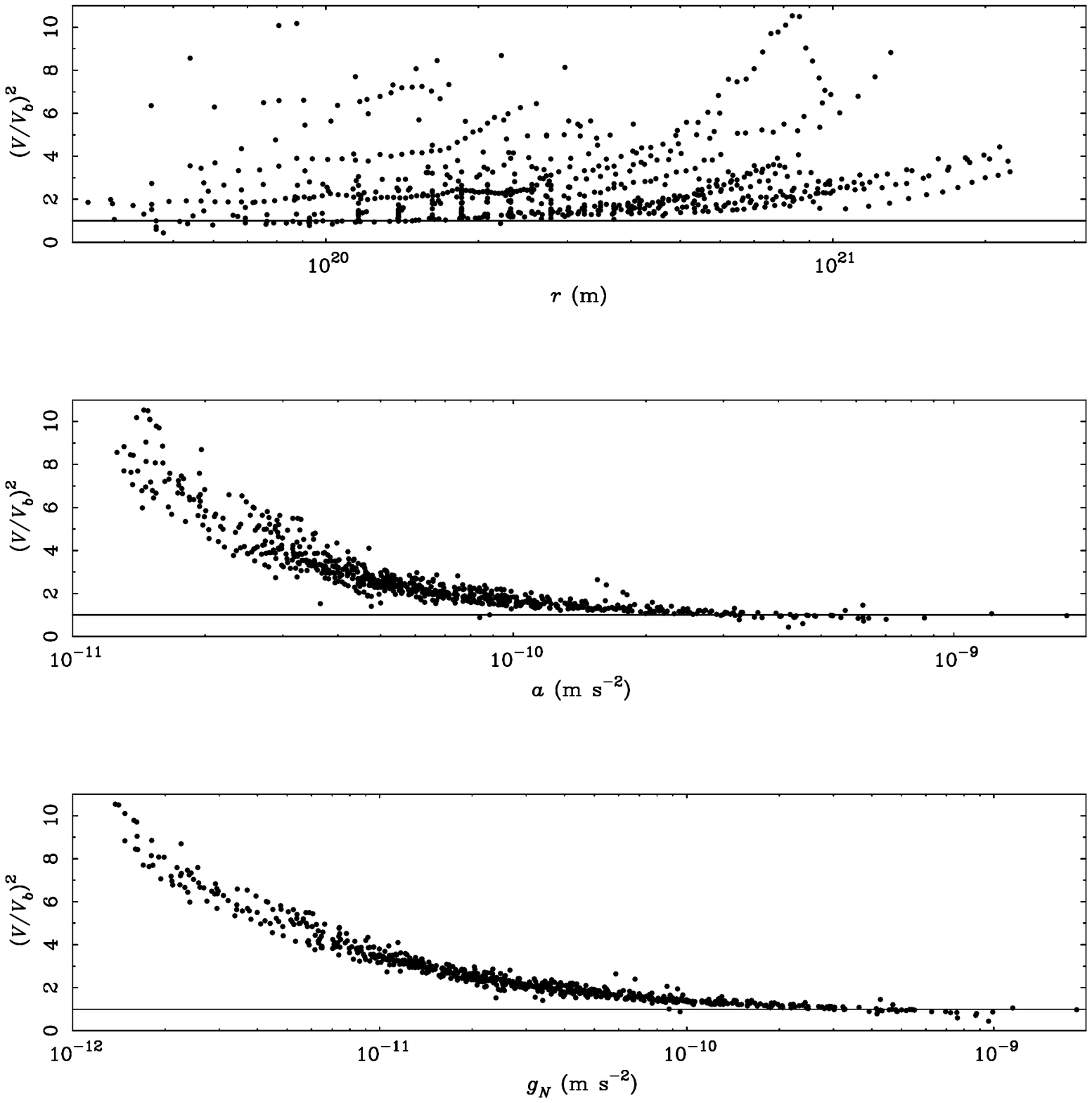}}
\vspace{-3.5cm} \caption{\footnotesize{The mass discrepancy, defined
    by the ratio $(V/V_\text{b})^2$ where $V$ is the observed velocity
    and $V_\text{b}$ is the velocity attributable to visible baryonic
    matter, in spiral galaxies. No correlation is found with the
    distance scale (top panel); however a strong correlation is seen
    with the acceleration scale (middle panel) and with the
    gravitational field scale (bottom panel). The mass discrepancy
    (\textit{i.e.} the presence of dark matter) appears below the
    critical acceleration scale $a_0\sim
    10^{-10}\,\text{m}/\text{s}^2$.}}
\label{fig:MD}
\end{figure}

The rotation curves of almost all spiral galaxies are reproduced in
great details with a single-parameter fit --- the mass-to-luminosity
ratio which is \textit{a posteriori} seen to be consistent with the
expectations coming from stellar populations. The baryonic
Tully-Fisher (BTF) relation~\cite{TF77,McG00}, an empirical relation
between the asymptotic rotation velocity and the baryonic mass of
galaxies, and valid for a large range of masses of
galaxies~\cite{FamMcG12}, is naturally reproduced. In particular, for
dwarf galaxies dominated by the gas there is little uncertainty on
both the rotation velocity and the baryonic mass, so the evidence for
the BTF relation is very strong~\cite{McG11}. The \textit{original
  sin} of MOND is Milgrom's law, namely that the discrepancy between
the dynamical and luminous masses, \textit{i.e.}  the presence of dark
matter, is correlated with the involved scale of acceleration or
magnitude of the gravitational field, see Fig.~\ref{fig:MD} which is
taken from Ref.~\cite{FamMcG12}.

In this paper we shall adopt for MOND the modified Poisson
equation\footnote{Boldface letters indicate ordinary Euclidean
  vectors; $G$ is Newton's gravitational constant.} of Bekenstein \&
Milgrom~\cite{BekM84}
\be{BMeq} \bm{\nabla}\cdot\biggl[\mu\left(\frac{g}{a_0}\right) \bm{g}
  \biggr] = - 4\pi G\,\rho_\text{b} \,, \ee
where $\rho_\text{b}$ denotes the ordinary mass density of the
baryons. The gravitational field is irrotational,
$\bm{g}=\bm{\nabla}U$, with $U$ the gravitational potential, and we
denote its norm by $g=\vert\bm{g}\vert$. The function $\mu$ of the
ratio $g/a_0$ is the MOND interpolating function, which interpolates
between the Newtonian regime $g\gg a_0$ for which $\mu\simeq 1$ (thus
one recovers the usual Poisson equation of Newtonian gravity in this
regime), and the MOND weak-field regime $g\ll a_0$ for which $\mu$ is
linear in its argument, $\mu\simeq g/a_0$. The constant $a_0$
represents the MOND acceleration scale separating the two regimes as
evidenced by Fig.~\ref{fig:MD}.

Several relativistic MOND theories, extending general relativity with
new fields and without the need of dark matter, have been proposed:
\begin{itemize}
\item The Tensor-Vector-Scalar (TeVeS) theory, which extends general
  relativity with a time-like vector field and one scalar
  field~\cite{Sand97,Bek04,Sand05};
\item Einstein-{\ae}ther theories, originally motivated by the
  phenomenology of Lorentz invariance violation~\cite{JM00,JM04},
  involve a unit time-like vector field non-minimally coupled to the
  metric and with a non-canonical kinetic term~\cite{ZFS07,Halle08};
\item A bimetric theory of gravity in which the two metrics are
  coupled through the difference of their Christoffel
  symbols~\cite{bimond1,bimond2};
\item A variant of TeVeS using a Galileon field and a Vainshtein
  mechanism to prevent deviations from general relativity at small
  distances~\cite{BDgef11};
\item A theory based on a preferred time foliation labelled by the
  so-called Khronon scalar field~\cite{BM11,Sand11}.
\end{itemize}
The cosmology of these theories has been extensively investigated,
notably in TeVeS and non-canonical Einstein-{\ae}ther
theories~\cite{Sk06,LiMo08,Sk08,Zu10}. However, all these theories
have difficulties in reproducing the CMB spectrum, even when adding a
component of hot dark matter~\cite{Sk06}.

\section{The Dielectric Analogy of MOND}
\label{sec:dielectric}

MOND enjoys a remarkable property, known as the dielectric analogy,
which could have far-reaching implications. Indeed the MOND
equation~\eqref{BMeq} represents exactly the gravitational analogue
(in the non-relativistic approximation) of the Gauss equation of
electrostatics when modified by polarization effects taking place in
non-linear dielectric media~\cite{B07mond}. Taking this analogy at
face, we can interpret the MOND function $\mu$ entering
Eq.~\eqref{BMeq} as a coefficient characterizing some
``digravitational medium'', and write it as
\begin{equation}\label{muchi}
\mu = 1 + \chi\,,
\end{equation}
where $\chi$ would represent the gravitational susceptibility of this
medium, parametrizing the relation between the polarization, say
$\bm{P}$, and the gravitational field,
\begin{equation}\label{Pg}
\bm{P} = - \frac{\chi}{4\pi G} \,\bm{g}\,.
\end{equation}
Thus $\chi$ characterizes the response of the digravitational medium
to an applied gravitational field. From Eq.~\eqref{BMeq} we see that
the susceptibility coefficient depends on the norm $g$ of the
gravitational field, in close analogy with the electrostatics of
non-linear media. The mass density associated with the polarization is
given by the same formula as in electrostatics,
\begin{equation}\label{rhopol}
\rho_\text{pol} = -\bm{\nabla}\cdot\bm{P}\,.
\end{equation}
With those notations Eq.~\eqref{BMeq} can be rewritten as
\begin{equation}\label{MONDpol}
\Delta U = -4\pi G \,\bigl(\rho+\rho_\mathrm{pol}\bigr)\,,
\end{equation}
indicating that the Newtonian law of gravity may not be violated, but
that we are facing a new form of dark matter, in the form of polarized
masses with density $\rho_\mathrm{pol}$.

Let us proceed further and view the dark matter medium as consisting
of individual dipole moments $\bm{p}$ with number density $n$, so that
the polarization reads
\begin{equation}\label{Pp}
\bm{P} = n\,\bm{p}\,.
\end{equation}
We suppose that the dipoles are made of a doublet of sub-particles,
one with positive gravitational mass $m_\text{g}=+m$ and one with
negative gravitational mass $m_\text{g}=-m$, in analogy with electric
charges. If the two masses are separated by the spatial vector
$\bm{\xi}$, pointing in the direction of the positive mass, the dipole
moment is
\begin{equation}\label{pxi}
\bm{p} = m\,\bm{\xi}\,.
\end{equation}
Let us further suppose, still with analogy with electric charges, that
the sub-particles have positive inertial masses $m_\text{i}=m$, so
that the dipole moment consists of an ordinary particle
$(m_\mathrm{i},m_\mathrm{g})=(m,m)$ associated with an exotic one
$(m_\mathrm{i},m_\mathrm{g})=(m,-m)$.

The ordinary particle will always be attracted by some mass
distribution made of ordinary matter, while the other particle
$(m_\mathrm{i},m_\mathrm{g})=(m,-m)$ will always be repelled by the
same mass distribution. In addition the two sub-particles would repel
each other. We see therefore that the gravitational dipole is
unstable, and we need to invoke a non-gravitational internal force to
supersede the gravitational force between the
sub-particles~\cite{B07mond}.

Simply from these considerations we expect that an external
gravitational field will exert a torque on the dipole moment in such a
way that its orientation will have the positive mass oriented in the
direction of the external mass, and the negative one oriented in the
opposite direction. Thus we find that $\bm{p}$ and $\bm{P}$ should
point towards the external mass, \textit{i.e.} be oriented in the same
direction as the gravitational field $\bm{g}$. From Eq.~\eqref{Pg} we
therefore conclude that
\begin{equation}\label{chisign}
\chi < 0\,.
\end{equation}
This corresponds to an ``anti-screening'' of ordinary masses by the
polarization masses, and an enhancement of the gravitational field in
the presence of the digravi\-tational medium. The
result~\eqref{chisign} is nicely compatible with the prediction of
MOND, since we have $\mu=1+\chi\sim g/a_0\ll 1$ in the MOND
regime. The phenomenology of MOND can thus be interpreted (at the
non-relativistic level) as resulting from an effect of gravitational
polarization, of some cosmic fluid made of polarizable dipole moments,
aligned with the gravitational field of ordinary matter (galaxies),
and representing a new form of dark matter.

Of course the previous interpretation of dark matter rings a bell, and
it is tempting to interpret this polarizable medium as a sea of
virtual pairs of particles and antiparticles. Although this idea poses
a lot of problems, let us examine a few orders of magnitude that such
an hypothetical medium would have. We thus suppose that the dark
matter medium is made of virtual particle-antiparticle pairs $(m,\,\pm
m)$, with polarisation field~\eqref{Pp} and individual dipole
moments~\eqref{pxi}. The classical separation between particles and
antiparticles should be of the order of magnitude of the Compton
wavelength, hence $\xi\sim\lambda_\text{C}\equiv \hbar/m\,c$, and thus
\be{P} P \sim \dfrac{n\,\hbar}{c} \,.  \ee
On the other hand, in the MOND regime we have $g\simeq a_0$, so from
Eq.~\eqref{Pg} with $\chi\simeq -1$ the polarisation field is of order
$P\simeq a_{0}/4\pi G$, hence we obtain the following estimation of
the medium density,
\be{n} n \sim \frac{a_{0}\,c}{4\pi\,G\,\hbar}\sim 4.3\times 10^{35}
\,\mathrm{cm}^{-3}\,, \ee
where we have adopted the common value of the MOND acceleration
$a_{0}=1.2\times 10^{-10} \,\mathrm{m/s}^2$. This gives a
characteristic length for the separation inside pairs,
\be{xi} \xi \sim n^{-1/3} \sim 1.3\times 10^{-12} \,\mathrm{cm} \,,
\ee
and an estimation of the mass of the dark matter particles,
\be{m} m\sim\dfrac{\hbar}{\xi\, c} \sim 14\, \mathrm{MeV} \,.  \ee

Interestingly, these estimations, in which the value of MOND's
acceleration scale $a_0$ plays the crucial role, turn out to be very
close to typical estimations for the standard QCD
vacuum~\cite{Chardin09,Hajdu11}. Regardless of this fact being a
coincidence or not, recall that here we made the wild assumption that
antiparticles have mass $(m_\mathrm{i},m_\mathrm{g})=(m,-m)$ which is
at odds with all theoretical expectations~\cite{MG57}. Furthermore
this assumption is severely constrained by equivalence principle
E\"otv\"os-type experiments using the virtual $e^+e^-$ and $q\bar{q}$
pairs in ordinary materials~\cite{schiff59}. Note also that the above
description of vacuum fluctuations based on Compton's separation is
merely semi-classical and probably oversimple.

\section{Dipolar Dark Matter and Modified Gravity}
\label{sec:DDM}

Some aspects of the previous model have been promoted to a
relativistic description in the concept of dipolar dark matter --- a
form of matter described by a relativistic current and endowed with a
space-like vector field called the dipole moment, and obeying a
specific Lagrangian in standard general relativity~\cite{BL08,BL09}.
But obviously, because of the negative masses, not all aspects of the
model could be made compatible with general relativity, in particular
it was impossible to give to the dipole moment a microscopic
interpretation in terms of sub-particles.

In the present section we shall point out that, in certain conditions,
it is possible to mimic the effect of gravitational polarization (and
the involved anti-gravity) by coupling the two species of
sub-particles to two different Newtonian potentials. We shall provide
a non-relativistic model and show how it recovers exactly the MOND
equation~\eqref{BMeq} in all non-spherical and dynamical
situations. Furthermore this new model will be amenable to a
relativistic extension based on a bimetric coupling of dark matter
particles~\cite{BB14}.

We consider the following non-relativistic Lagrangian for the dynamics
of matter fields, consisting of ordinary baryons and two species of
dark matter particles, and coupled to gravity:
\begin{eqnarray}
L &=& \int\ud^{3}\bm{x} \left\{
-\frac{1}{\kappa}\vert\bm{\nabla}U\vert^2
-\frac{1}{\kappa}\vert\bm{\nabla}\ul{U}\vert^2 -
\frac{1}{2\varepsilon}\vert\bm{\nabla}(U+\ul{U})\vert^2\right. \nonumber\\ &&
\qquad\quad~\left. +
\rho_\text{b}\Bigl(U+\frac{\bm{v}^2_\text{b}}{2}\Bigr) +
\rho\Bigl(U+\phi+\frac{\bm{v}^2}{2}\Bigr)\right. \nonumber\\ &&
\qquad\quad~\left. +
\ul{\rho}\Bigl(\ul{U}-\phi+\frac{\ul{\bm{v}}^2}{2}\Bigr) +
\frac{a_0^2}{2\alpha}\, W\left(X\right)\right\}\,.\label{lagrangien}
\end{eqnarray}
Here $\kappa$, $\varepsilon$ and $\alpha$ denote some coupling
constants to be specified later, and $a_0$ is the MOND acceleration
constant scale. The matter fields are described by their usual
Newtonian mass density and velocity: $(\rho_\text{b},
\bm{v}_\text{b})$ for the baryons, $(\rho, \bm{v})$ and $(\ul{\rho},
\ul{\bm{v}})$ for respectively the first and second types of dark
matter particles. These variables are linked by the ordinary
continuity equation, \textit{e.g.} $\partial_t\rho +
\bm{\nabla}\cdot(\rho\bm{v})=0$. In this model, the main point is that
the particles $(\rho, \bm{v})$ are coupled to the ordinary Newtonian
potential $U$ as for the baryons, but that the particles $(\ul{\rho},
\ul{\bm{v}})$ are coupled to a different potential $\ul{U}$. The two
potentials $U$ and $\ul{U}$ interact with each other in the way
specified by their kinetic terms in~\eqref{lagrangien}.

As in the model of Ref.~\cite{B07mond}, we need to introduce an
internal force to stabilize the dipole moment. This is described here
by a scalar potential $\phi$ obeying a non-canonical kinetic term
given by the last term in~\eqref{lagrangien}, which involves a
function $W$ of the ratio
\be{X} X\equiv\frac{\vert\bm{\nabla}\phi\vert^2}{a_{0}^{2}}\,.\ee
This function is determined phenomenologically so as to recover the
MOND phenomenology, but in principle it should be derived from some
more fundamental theory. In the limit when $X\to 0$, which will
correspond to the MOND regime, we impose
\be{W0} W = X-\frac{2}{3} X^{3/2} + \calO\left(X^2\right)\,.\ee
On the other hand, in order to recover the Newtonian limit, it will be
sufficient to impose that $W'\equiv\ud W/\ud X$ tends to zero when $X
\to +\infty$. We can already note that a stronger condition when $X
\to +\infty$, namely
\be{Winf} W = A + \frac{B}{X} + \calO\left(\frac{1}{X^2}\right)\,,\ee
where $A$ and $B$ are some constants, will actually be better in order
to suppress all polarization effects in the Newtonian regime.

We now vary the Lagrangian with respect to all particles and
fields. The equation of motion of baryons is standard,
\be{eomNRb} \frac{\ud \bm{v}_\text{b}}{\ud t} = \bm{\nabla} U\,.\ee
At the contrary, because of the postulated internal potential
interaction $\phi$, we obtain for the dark matter particles,
\bes{eomNR} \frac{\ud \bm{v}}{\ud t} &=&
\bm{\nabla}\bigl(U+\phi\bigr)\,,\\\frac{\ud \ul{\bm{v}}}{\ud t} &=&
\bm{\nabla}\bigl(\ul{U}-\phi\bigr)\,.\ees
Varying with respect to $\phi$ we get
\be{eqgauss} \bm{\nabla}\cdot\Bigl[W'(X)\,\bm{\nabla}\phi \Bigr] =
\alpha \left(\rho - \ul{\rho} \right)\,.\ee
Finally, varying with respect to $U$ and $\ul{U}$ we get two
equations, which can conveniently be re-arranged into
\begin{subequations}\label{eqUUbar}
\begin{align} &\Delta U = - \frac{\kappa^2}{4(\kappa+\varepsilon)}
\biggl[\Bigl(1+\frac{2\varepsilon}{\kappa}\Bigr)
  \bigl(\rho_\text{b}+\rho\bigr)-\ul{\rho}\biggr]\,,\\ &\Delta\bigl(U+\ul{U}\bigr)
= - \frac{\kappa\varepsilon}{2(\kappa+\varepsilon)}
\bigl(\rho_\text{b}+\rho+\ul{\rho}\bigr)\,.\end{align}\end{subequations}

The condition under which our model will work, \textit{i.e.} where a
mechanism of gravitational polarization will show up, is
\be{epskappa} \varepsilon \ll \kappa\,.\ee
As is already seen at the level of the Lagrangian~\eqref{lagrangien},
such a condition in the coupling constants forces the two potentials
$U$ and $\ul{U}$ to be (approximately) opposite to each
other. Therefore, under this condition, we obtain the following
Poisson equation for the ordinary Newtonian potential $U$ felt by the
baryonic matter,
\be{Uord} \Delta U = - \frac{\kappa}{4}\bigl(\rho_\text{b}+\rho -
\ul{\rho}\bigr) \,,\ee
the potential in the other sector being given by $\ul{U} = -U$.

We now look for a plasma-like solution of these equations. For this
purpose, we assume the existence of an equilibrium configuration with
uniform density $\rho_0$, and that the two dark matter fluids are
displaced with respect to this equilibrium. Their densities can thus
be written as
\bes{rhosol} \rho &=& \rho_0 -
\frac{1}{2}\,\bm{\nabla}\cdot\bm{P}\,,\\ \ul{\rho} &=& \rho_0 +
\frac{1}{2}\,\bm{\nabla}\cdot\bm{P}\,.\ees
Here we defined the polarization $\bm{P}=\rho_0\,\bm{\xi}$ where
$\bm{\xi}$ denotes the Eulerian relative displacement; thus, $\rho_0=n
m$ in the notation of
Eqs.~\eqref{Pp}--\eqref{pxi}. Using~\eqref{rhosol} we can solve for
the internal field equation~\eqref{eqgauss},
\be{Pphi} \alpha\,\bm{P} = - W'\,\bm{\nabla}\phi \,,\ee
which shows that the polarization is aligned with the internal
field. However, it is not that obvious that it will also be aligned
with the gravitational field (``gravitational polarization''). This
will come from the equations of motion of the dark matter particles
which now read (since $\ul{U}=-U$)
\bes{eomsol} \frac{\ud \bm{v}}{\ud t} &=&
\bm{\nabla}\bigl(U+\phi\bigr)\,,\\\frac{\ud \ul{\bm{v}}}{\ud t} &=& -
\bm{\nabla}\bigl(U+\phi\bigr)\,.\ees
As we see with this mechanism, the effective ratio between the
gravitational mass and the inertial mass of these particles appears to
be $m_\text{g}/m_\text{i}=\pm 1$, in agreement with the picture
proposed in Sec.~\ref{sec:dielectric}. However, with this new
description the non-relativistic Lagrangian can be generalized to a
relativistic formulation~\cite{BB14}.

Considering now the relative acceleration combined with
Eq.~\eqref{Pphi}, we obtain an harmonic oscillator for the
polarization $\bm{P}$ (or equivalently the displacement $\bm{\xi}$)
embedded in the gravitational field $\bm{g}=\bm{\nabla}U$,
\be{harmosc} \frac{\ud^2 \bm{P}}{\ud t^2} + \omega_0^2 \bm{P} = 2
\rho_0\,\bm{g}\,.\ee
The dark matter medium undergoes oscillations with plasma
frequency\footnote{Of course this is analogous to the classic
  derivation of the plasma frequency, see \textit{e.g.}
  Ref.~\cite{jackson}.}
\be{plasma} \omega_0 = \sqrt{\frac{2\alpha\,\rho_0}{W'}}\,.\ee
This imposes the coupling constant $\alpha$ to be positive. Note that
from Eq.~\eqref{W0} we have $W'=1+\mathcal{O}(X^{1/2})$ which is
positive in the MOND regime. Finally, averaging over the plasma
oscillations we obtain that the polarization is indeed aligned with
the local gravitational field, \textit{i.e.} $\bm{P} =
2\rho_0\,\bm{g}/\omega_0^2$, or equivalently
\be{Paligned} \bm{P} = \frac{W'}{\alpha}\,\bm{g}\,.\ee
Comparing with~\eqref{Pphi} we see that this simply means that
$\bm{\nabla}\phi=-\bm{g}$, which can also be deduced from
Eqs.~\eqref{eomsol} when the particles are in average at rest.

Finally the MOND equation follows immediately from the Poisson
equation~\eqref{Uord}. By inserting Eqs.~\eqref{rhosol} in
Eq.~\eqref{Uord} we transform it into
\be{Upol} \bm{\nabla}\cdot\Bigl[\bm{g} - \frac{\kappa}{4} \bm{P}\Bigr]
= - \frac{\kappa}{4}\,\rho_\text{b} \,,\ee
which really looks like an ordinary Poisson equation modified by
polarization effects. With the constitutive relation~\eqref{Paligned}
we recover the Bekenstein \& Milgrom~\cite{BekM84} form,
\be{mondeq} \bm{\nabla}\cdot\Bigl[\mu\Bigl(\frac{g}{a_0}\Bigr)\,\bm{g}
  \Bigr] = - \frac{\kappa}{4}\,\rho_\text{b} \,,\ee
with MOND interpolating function $\mu = 1 - \kappa\,W'/4\alpha$. It is
then easy to see that with the postulated form of the function $W$ in
Eq.~\eqref{W0} and the following values of the coupling constants:
\be{kappaalpha} \kappa = 4\alpha = 16\pi G\,, \qquad
\varepsilon\ll\kappa \,,\ee
we recover exactly the MOND regime when $g\ll a_0$, \textit{i.e.}
\be{muMOND} \mu \equiv 1 - W' = \frac{g}{a_0} +
\calO\left(\frac{g^2}{a_0^2}\right)\,.\ee
Thus the phenomenology of MOND appears to be a natural prediction of
this model. To recover the ordinary Poisson equation in the Newtonian
regime it suffices that $W'$ tends to zero when $X\to\infty$ (where
now $X=g^{2}/a_{0}^{2}$). However there may still be a residual
polarization in this limit, see Eq.~\eqref{Paligned}. As already
mentionned, to suppress it we prefer to impose the stronger condition
that $X W'$ tends to zero when $X\to\infty$, for instance the
behaviour given by Eq.~\eqref{Winf}.

\bibliographystyle{ws-ijmpcs}
\bibliography{wag_arxiv}

\end{document}